# Recoilless Resonant Capture of Antineutrinos from Tritium Decay

R. S. Raghavan

*Institute of Particle, Nuclear & Astronomical Sciences and Department of Physics*
*Virginia Polytechnic Institute and State University, Blacksburg VA 24061*

Antineutrinos ($\tilde{\nu}_e$) from bound state β-decay of tritium (T) can be resonantly captured in $^3$He. With T and $^3$He embedded in metal tritides, *recoilless* resonant capture may be possible with ultra sharp ($\Delta E/E \sim 4.5 \times 10^{-16}$) 18.6 keV $\tilde{\nu}_e$'s from T with an effective σ(res) $\sim 3 \times 10^{-33}$ cm$^2$. A specific technical approach that confronts stringent practical constraints for realizing recoilless $\tilde{\nu}_e$ is described. With the low $\tilde{\nu}_e$ energy and high σ(res), the $\tilde{\nu}_e$ gravitational shift, $\tilde{\nu}_e$ conversion to sterile ν and eventually, $\theta_{13}$ $\tilde{\nu}_e$ oscillations could be studied, all with g-quantities of material in bench scale baselines.

Recoilless neutrino reactions, first considered 45 years ago,[1] remain yet speculative because of the very stringent, unanswered experimental demands. State-of-the-art hydrogen storage technology now suggests new paths for progress. In this paper I outline a specific approach to observe recoilless resonant capture of 18.6 keV antineutrinos ($\tilde{\nu}_e$) emitted in the bound state β-decay of tritium (T), in a $^3$He target. The $\tilde{\nu}_e$ energy width $\Delta E/E \sim 4.5 \times 10^{-16}$ possible in this method, enhances the effective $\tilde{\nu}_e$ cross section σ(res) by $\sim 3 \times 10^9$ over the usual σ($\tilde{\nu}_e$+p). That opens new horizons for $\tilde{\nu}_e$ research with *grams* (not *kton*s) of material and (due to the low $\tilde{\nu}_e$ energy), baselines in the *bench scale* (not kms). Laboratory studies of the $\tilde{\nu}_e$ gravitational shift, $\tilde{\nu}_e \to$ sterile ν conversion[2] and $\theta_{13}$ $\tilde{\nu}_e$ oscillations may then be possible.

Mikaelyan et al[3] first noted the $\tilde{\nu}_e$ reaction

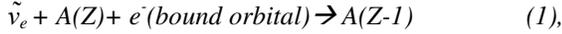

$$\tilde{\nu}_e + A(Z) + e^-(\text{bound orbital}) \to A(Z-1) \qquad (1),$$

where the bound electron is captured in the target atom A(Z) and stressed its resonant character at the energy:

$$E(\tilde{\nu}_{e\,res}) = Q + B_Z + E_R \qquad (2).$$

Q is the maximum $\tilde{\nu}_e$ energy (= $M_{Z-1} - M_Z$) in the β-decay A(Z-1)$\to$A(Z), $B_Z$, the binding energy of the electron in (1) from (the lowest) orbit in the nucleus A(Z) and $E_R$ the nuclear recoil energy. The reaction rate of (1) is[3] $R \propto |\Psi^2| \rho(E_{\tilde{\nu}e\,res}) / ft_{1/2}$. $ft_{1/2}$ is the reduced half-life of the reverse decay A(Z-1)$\to$A(Z), $|\Psi^2|$, the orbital electron density in the nucleus A(Z) and $\rho(E_{\tilde{\nu}e\,res})$, the resonant $\tilde{\nu}_e$ density, i.e., the number of $\tilde{\nu}_e$/unit energy interval at $E_{\tilde{\nu}e\,res}$. Reaction (1) is not useful with wide band $\tilde{\nu}_e$ from nuclear reactors[3] since $\rho(E_{\tilde{\nu}e\,res})$ is too small, but it can be greatly enhanced using the decay:

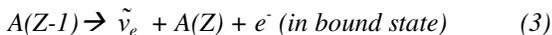

$$A(Z-1) \to \tilde{\nu}_e + A(Z) + e^- \text{ (in bound state)} \qquad (3)$$

that emits a *monoenergetic* $\tilde{\nu}_e$ with just the resonance energy (2). The electron emitted in the β-decay (3) of A(Z-1) is inserted in a vacant orbit in A(Z) instead of going into the continuum (Cβ). This process is called bound state β-decay (Bβ)[4]. The 2-body decay (3) results in a monoenergetic $\tilde{\nu}_e$ with $E_{\tilde{\nu}e} = Q + B_Z - E_R$ where $B_Z$ is *gained* in inserting an electron in the same orbit as in the capturing target A(Z) in (1). This energy is just the resonance energy (2) except for a deficit of $2E_R$ due to nuclear recoil in emission and capture. If this deficit is removed, the condition for resonant $\tilde{\nu}_e$ capture is precisely satisfied. The rate R is set by $\rho(E_{\tilde{\nu}e\,res})$ i.e., the *line width Γ* of the $\tilde{\nu}_e$ from the Bβ-decay (3).

The T-$^3$He system is ideal for observing resonant $\tilde{\nu}_e$ capture. It offers a super-allowed β-decay ($ft_{1/2}$ =1132 s), a very low energy $\tilde{\nu}_e$ (18.6 keV) and a sizable Bβ branching[4] ($\sim 5.4 \times 10^{-3}$) to the atomic ground state of $^3$He. The initial T atom has a vacancy in the 1s shell for Bβ decay and the target $^3$He has two 1s electrons one of which can be captured. 1s electron capture is ideal since $|\Psi^2|$(He n =1s) (for Bβ decay and $\tilde{\nu}_e$ capture) is $\propto 1/n^3$ and maximal for n=1.

The resonant $\tilde{\nu}_e$ capture cross section is[3]:

$$\sigma = 4.18 \times 10^{-41} g_o^2 \rho(E_{\tilde{\nu}e\,res}/\text{MeV}) / ft_{1/2} \; \text{cm}^2 \qquad (4)$$

with $g_o^2 \sim 1.24 \times 10^{-5}$ for 1s electrons. Consider gaseous T and $^3$He at 300 K. In this case the $\tilde{\nu}_e$ energy has a thermal Doppler profile with a width $2\Delta \sim 2E_\nu(2kT/Mc^2)^{1/2} = 0.16\; eV$ (k = Boltzmann constant). Thus ρ in (4) is enhanced by $\sim 10^6/0.16 \sim 6 \times 10^6$. Since $2\Delta \sim 2E_R = 2E_\nu^2/Mc^2 = 0.123\; eV$ the total recoil energy shift, the incident $\tilde{\nu}_e$ shape overlaps ($\sim 0.4$) with the absorption window. Thus, σ $\sim 10^{-42}$ cm$^2$, similar to the usual $\tilde{\nu}_e + p \to n + e^+$ reaction (at $E_\nu$=3 MeV), even though $E(\tilde{\nu}_e)$ is 150 times smaller. Indeed, σ($\tilde{\nu}_e$+p) = 0 for $E(\tilde{\nu}_e)$ <1.8 MeV. The prospect of detecting 18.6 keV $\tilde{\nu}_e$'s for the first time is attractive, e.g., for $\theta_{13}$ $\tilde{\nu}_e$ oscillation experiments that need a baseline L(m)$\sim 0.5E$(keV). For this case, only 9.3 m is needed vs. >1.5 km for 3MeV reactor $\tilde{\nu}_e$'s. However, the T$\to ^3$He thermal resonance rate at 9.3 m is only 0.03/day for a 100 MCi T source and 1kg $^3$He.

Is it possible to increase σ in (4) by enhancing the resonant density $\rho(E\tilde{\nu}_{eres})$ further? The question leads logically to *recoilless* $\tilde{\nu}_e$ emission and capture (Mőssbauer Effect)[5] in the $T\beta\rightarrow{}^3He$ system that eliminates the resonance mismatch due to recoil. If the source and target atoms are embedded in *solids*, lattice dynamics suppress the $\tilde{\nu}_e$ recoils in a fraction f~ exp–[(4/3 ($E_R$/Z)] where Z is the zero point energy (ZPE) of emitter/target. An effective $\tilde{\nu}_e$ line width $\Gamma <<\Delta$ can then enhance the $\tilde{\nu}_e$ resonant density beyond thermal resonance values.

The T→He $\tilde{\nu}_e$ emission and absorption are precisely time reversed processes that naturally ensure exact energy conservation. In the $\tilde{\nu}_e$ emission, the initial state is T which is chemically bound in a solid. In the final state a neutral He +$\tilde{\nu}_e$ is created by Bβ-decay with no phonon emission. In $\tilde{\nu}_e$ absorption, the initial state is $\tilde{\nu}_e$+neutral He and the final state after electron capture, a (bound) T and no phonons. The $\tilde{\nu}_e$ energy is, in general, modified by energies $E_T$ and $E_{He}$ (which include atomic and chemical binding energies, the lattice vibration energy including the ZPE and the rigid lattice dipolar interaction energy). If the perturbing energies $E_T$ and $E_{He}$ are *unique, static, and identical* in source/absorber, a deficit ($E_T$-$E_{He}$) in the Bβ transmutation T→He at $\tilde{\nu}_e$ emission is self-compensated *exactly* in reverse by ($E_{He}$-$E_T$) in the He→T e-capture transmutation in absorption. Violations of this rule can be expected, e.g., via unequal temperatures of the source/absorber that result in a shift $\Delta E/E= (3/2\ k\Delta T/\ Mc^2)$ due to the second order Doppler effect. Identical cryogenic baths for source/absorber are thus mandatory. Even if the T and He sites are unique, in principle the ZPE could vary slightly from site to site, causing inhomogeneous line broadening. The $\tilde{\nu}_e$ line is also broadened homogeneously by dipolar relaxation. The resonance effect is diluted by the largest of the line broadening types. The T and $^3$He have zero quadrupole moment (spin ½) that eliminates the entire class of electric perturbations. The T-$^3$He matrix should thus be designed for *unique* and *identical* sites for T and He each in the source/absorber.

Metal tritides[6] offer a practical approach for embedding T and $^3$He in solids. Hydrogen (T) chemically reacts with metals to form hydrides (tritides). The process ensures uniform population of T in the bulk of the metal. As the tritide ages, the $^3$He daughter grows and populates the lattice uniformly ("the tritium-trick" TT). However, the parent T still in the target must be removed (see below) in order to observe the resonance.

Can the TT method provide unique and identical T and He sites in the source and target? The site of the source He is its birth site –that of its parent T. The site of He in the absorber is the final site of a mobile inert atom with possibly different and indeed, non-unique site types. The T and He sites in the TT approach have been studied extensively. In bcc metals (Ti, Nb, V), the T sits exclusively in tetragonal interstitial sites (TIS) whereas in fcc metals (Pd, Ni), it finds octahedral interstitial sites (OIS)[7]. The case of He is more complex. Because of its insolubility it is highly mobile and thus could form clusters and eventually, micro-bubbles of typical radius ~2nm containing ~4000 atoms. At temperatures below ~100K the bubble He is solid under high pressure (~10 GPa). The bubble sizes and pressures are not unique. Even in the best cases (TaT, VT) the He atomic volumes in the bubbles and thus the ZPE's vary by ~2-5%[8] that leads to $\Delta E/E$ of 1000's of line widths. Thus TT conditions that lead to He bubbles are basically unsuitable for recoilless $\tilde{\nu}_e$ resonance.

Is it possible to engineer TT conditions to avoid He bubbles? The parameters of the problem are: the He diffusivity D(K) at temperature K, the He generation rate g =(T/Metal M)1.79x10$^{-9}$/s, and the activation energies E1, E2 and E3 for jumps, pair cluster formation and bubble coalescence (see Table 1). A set of coupled non-linear differential equations[9] describe the time evolution of the concentrations c1 (mobile interstitials), c2 (pair clusters) and c3 (bubbles). These equations were solved numerically[10] for NbT and TiT with similar results. For PdT the results agreed with experimental data. The results for NbT are shown in Fig. 1. The graphs show the growth of He for 200 days when the T is switched off by desorption. Thereafter, the He in the T-free sample has different ratios of IS/(bubbles) = c1/(c2+c3) at different temperatures, exemplified by a flat c1 (as at 200K) or a decaying c1 indicating loss to growing bubbles c2+c3, shown explicitly in the lower curves for T>235K. The 200K results in Fig. 1 are not very sensitive to the exact parameter values: x100 larger D and/or a smaller E1 =

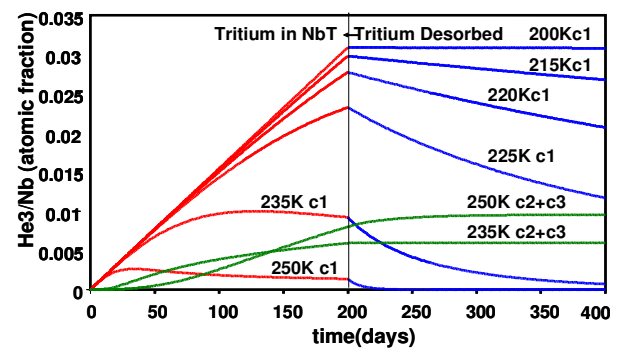

Fig 1 $^3$He is generated in $Nb_{1.0}T_{1.0}$ for 200 days, when the He generating T is desorbed. The figure shows the $^3$He concentration in interstitial sites (c1)(t) and that in clusters/bubbles (c2+c3) (t) for different ambient temperatures and time t, assuming the He mobility/activation parameters in Table 1. The He in the T-free absorber below 220K is almost all interstitial and above 235K, almost all in clusters/ bubbles.

0.8 eV do not change the results. Thus, in NbT the $^3$He reside *only* at unique IS sites if T <200K, grows indefinitely and remains without bubble formation after the T is removed. This behavior in NbT (and TiT) is exceptional. In most other tritides (e.g. PdT), bubble formation dominates already at >20K.

In the bcc NbT source, the T and thus, also the just-born daughter He reside in the TIS[7] (see Table 2 for the Self-trapping energy EST at TIS and OIS). In the target the He would normally reside only in OIS (bcc or fcc). However, *Nb is unique* with degenerate EST(TIS) and EST(OIS) for He (Table 2). Thus, both sites can be randomly filled with equal probability. There are 6 TIS and 3 OIS in the bcc unit cell. The ZPE at the two sites (Table 2) differ by ~0.01eV so that the 33% He in OIS are off resonant (by ~$10^4$ fractional line widths), a small cost. Thus, NbT aged below 200K offers unique, identical TIS for the emitter T-He and (most of) the absorber He (the sites of H(T) in Nb have been verified by ion channeling techniques.[11]) In contrast, in TiT where bubbles are also avoided as in NbT, the emitter T is in a TIS but the absorber He sits *only* in the off-resonant OIS. The *NbT case is thus a key discovery* of at least one case that meets in a single strategy, the diverse, stringent demands on a viable T-He matrix.

Using the ZPE data in Table 2 for He and T in Nb(TIS) and ZPE = Z, the recoil free fraction f = f(He)f(T)≈ exp–{(27$E_R$/16) [1/Z(T)+1/Z(He)]}≈ 0.076. The homogeneous $\tilde{\nu}_e$ line width in NbT arises from dipolar relaxation with neighbor nuclear spins of $^3$He, T and $^{93}$Nb and the host paramagnetic susceptibility. A relaxation line width of ~13 kHz measured at H in NbH[12] closely applies to the case of NbT. This leads to a $\tilde{\nu}_e$ line width $\Gamma$~ 8.6x$10^{-12}$ eV ($\Delta E/E$ ~4.6x$10^{-16}$).

The unknown factor is inhomogeneous broadening due to possible site variation of the ZPE. The experimental observation of the ultra sharp Mőssbauer γ-resonance in $^{67}$Zn (natural width $\Delta E/E$ ~5x$10^{-16}$, very similar to the T-He $\tilde{\nu}_e$ resonance width above) and its gravitational shift (of $\Delta E/E$ <$10^{-17}$)[13], offers a direct precedent. In $^{67}$Zn, not only non-unique ZPE but also inhomogeneous electric fields of lattice strains and defects come into play via the non-zero quadrupole moment of $^{67}$Zn. The positive results *despite* these additional perturbations suggest that the inhomogeneous $\tilde{\nu}_e$ line width in the tritides may be reasonably comparable to the homogeneous width estimated above.

The basic design emerging from the above discussion is a source and target of NbT aged in an identical manner at <200K. This still leaves the problem of removing the T in the target (by a factor ~$10^{17}$). A possible approach to remove the T is isotopic exchange T→D (the exchange energies are very small: e.g., $E_{ex}$(D→ H; = 0.043 eV[14], even smaller for T→D or H).

*Table 1 He transport parameters in NbT at 200K*

| $M_1T_1$ | E1 eV | E2 eV | E3 eV | D/cm$^2$ s |
|---|---|---|---|---|
| M=Nb | 0.9[a] | 0.13[b] | 0.43[b] | 1.1E-26[c] |

[a] Ref. 7; [b] Ref. 9; [c] Assumes *tritium* pre-exponential $D_0$ (ref. 6)

*Table 2. Theoretical (Ref. 7) EST & ZPE for T and $^3$He in Nb interstitial sites (IS)*

| Site | EST (eV) | | ZPE (eV) | |
|---|---|---|---|---|
| | T | He | T | He |
| TIS | -0.133 | -0.906 | 0.071 | 0.093 |
| OIS | -0.113 | -0.903 | 0.063 | 0.082 |

*Table 3. Nearest neighbor (NN) Displacements(%) and measured[6] activation energies Eac(eV) in NbIS (Ref. 7)*

| | 1$^{st}$ NN Displacement | | | 2$^{nd}$ NN Displacement. | | |
|---|---|---|---|---|---|---|
| | H | D | T | H | D | T |
| TIS | 4.1 | 3.9 | 3.9 | -0.37 | -0.36 | -0.35 |
| OIS | 7.7 | 7.5 | 7.4 | 0.2 | 0.19 | 0.19 |
| Eac[6] | 0.106 | 0.127 | 0.135 | | | |

The limit of T exchange is set basically by the dilution factor so that repeated exchange cycles can be used. The method has been demonstrated on PdT$_{0.1}$[15] that was exposed to successive batches of ~35 Torr D$_2$ gas until mass spectrometry showed "nearly complete" absence of T (or presence of He) in the exchange gas. The method must be quantified anew by ultra-sensitive mass spectrometry. Methods to avoid T altogether by building metal layers (at 200K) with buried He are under study.

The isotopic exchange T→D introduces an ostensible difference in source NbT(He) and absorber NbD(He). Does this change the He ZPE and thus $E_{He}$-$E_T$ in the absorber? Table 3 shows that the % atomic displacements of 1$^{st}$ and 2$^{nd}$ nearest neighbors (NN) are hardly changed by the T→D exchange. The measured activation energies Eac (Table 3) are nearly the same for D and T (i.e. not ∝ 1/√M). It is thus likely that the T→D exchange makes only a small change if at all, on the He ZPE that can be compensated by a Doppler drive. The data suggest less preferable larger changes for T→H exchange that allows, however, a more efficient exchange T→H. Detailed data are needed for a judicious choice.

The T→D (or H) exchange also enables observation of the resonance signal by "milking" the created T and measuring its accumulation rate. The reverse β-activity Rβ of the final T created in the target is also a resonance signal. As the $\tilde{\nu}_e$ activation proceeds, Rβ increases (nearly) linearly with time, offering a signature. After a growth period Δt, Rβ(Δt)/day = (Captures/day) (Δt/6500d (= τ(T)). The basic background is the n+$^3$He→ T+p reaction. The $^3$He target must therefore be shielded against ambient neutrons at all times after removal of T.

A *real-time, T-specific* signal of the $\tilde{\nu}_e$ resonance in the target is of great interest. Rudimentary ideas for this purpose are suggested by: 1) A sudden change of the magnetic moment from -2.1 nm ($^3$He) → +2.79 nm (T) in the He→T transition This generates a transient (~ 0.1 ms) magnetic field that couples to the electron moment created in the final T via the hyperfine interaction. The response of the electron moment may be detectable by ultra-sensitive SQUID magnetometers. And 2) new electrons appear in the Nb d-band from the resonance T. They create additional specific heat that grows linearly with $\tilde{\nu}_e$ activation and possibly detectable by ultra-sensitive bolometric/micro-calorimetric techniques.

With $f$ ~0.076, $\rho(E_{\tilde{\nu}e\,res}/MeV) = 10^6/[\Delta E(res) = 8.56 \times 10^{-12}]$ and a He TIS site fraction of 2/3, the effective cross section (from (4)) for recoilless resonant capture $\sigma(res)$ is ~$0.3 \times 10^{-32}$ cm$^2$, ~$3 \times 10^9$ times larger than $\sigma(\tilde{\nu}_e+p)$ (3 MeV). Table 4 lists the capture rates and R$\beta$ signal rates (after an activation time $\Delta t = 65d = 0.01\,\tau(T)$) in 5cm and 10m baselines. Capture rates at close geometries are high enough (~0.05Hz) to provide safety margins against signal losses via unanticipated line broadening. In practice, 1 kCi T (= ~100mg T) needs 3g of Nb$_{1.0}$T$_{1.0}$ (note that ~0.5 kCi T in 2.4 g PdH$_{0.6}$ was used for NMR studies at low temperatures[16]). A 100mg target of $^3$He needs ~100g of Nb$_{1.0}$T$_{1.0}$ aged for 200 d.

*Table 4. T-$^3$He recoilless resonant capture rates*

| Base Line | T | $^3$He | $\tilde{\nu}_e$Capture/ d =Tmilked/ d | R$\beta$($\Delta t$ =65d)/ d |
|---|---|---|---|---|
| 5cm | 1kCi | 100mg | ~40x10$^3$ | ~40 |
| 10m | 1MCi | 1g | ~10$^3$ | ~10 |

Applications of recoilless $\tilde{\nu}_e$ resonance include general physics and neutrino physics. The priority of initial experiments at close geometries with high rates is to define the technical parameters for observing recoilless $\tilde{\nu}_e$ resonance. Important physics experiments may also be possible even in this phase. The Equivalence Principle predicts a shift (yet untested directly for neutrinos) of $\Delta E/E$ ~$g/c^2$ ~$10^{-18}$/cm fall, i.e. ~1 full line width (as obtained above) /400 cm fall. This effect may thus be accessible in bench scale fall heights. The important question of sterile neutrinos (of high current interest) could be investigated. The results of the LSND experiment,[17] are consistent with $\tilde{\nu}_e$ (mass)$^2$ difference $\Delta m^2$ ~1eV$^2$ (= ~$10^4$ x $\Delta m_{12}^2$ of reactor $\tilde{\nu}_e$) and mixing angle sin$^2$ 2$\theta$ ~0.1 to 0.001 that may imply conversion of $\tilde{\nu}_e$ → sterile $\nu$. It has been pointed out[2] that this result can be tested by reactor $\tilde{\nu}_e$ in short baselines that are, however, difficult to arrange in practice. For a 18.6 keV $\tilde{\nu}_e$, this implies ultra-short baselines as small as ~6 cm ($10^5$ m (in Kamland[18]) x$10^{-4}$ x18.6/3000). This key test for sterile neutrinos ideally fits the program at close geometries. The initial phase will clarify the feasibility of ~10 m base-lines for active $\tilde{\nu}_e$ $\theta_{13}$ flavor oscillations.

In summary, I have outlined a specific road map for the discovery of recoilless resonant $\tilde{\nu}_e$ capture based on the unique properties of $^3$He in NbT. The technique foresees dramatically new perspectives for contemporary neutrino physics research. New applications in other fields are likely with recoilless neutrinos as was the case historically with recoilless $\gamma$-rays.

This paper is dedicated to the memory of John N. Bahcall. I thank Walter Potzel (Technical University Munich) for the valuable critique that benefited this work. I thank Donald Cowgill and his colleagues at Sandia National Laboratory, Livermore for technical information on tritides.

[1] W. Kells and J .Schiffer, Phys. Rev. **C28** (1983) 2162; M.Visscher Phys. Rev. **116** (1959) 1581
[2] L. Mikaelyan et al, hep-ph/0310246.
[3] L. Mikaelyan et al, Sov J. Nucl. Phys **6** (1968) 254
[4] J. N. Bahcall, Phys. Rev. **124** (1961) 495
[5] H.Frauenfelder,*The Mössbauer Effect (*Benjamin*,* 1962)
[6] R. Lässer, *Tritium and $^3$He in Metals* (Springer, 1989)
[7] M. J. Puska & R. M. Nieminen, Phys. Rev. **B29** (1984) 5382
[8] T. Shober et al, Phys. Rev.**B40** (1989) 1277
[9] D. F. Cowgill, Sandia Natl. Lab. Report 2004-1739 (2004)
[10] I thank Prof. Kunghwa Park and Mr. D. Rountree for their help in these calculations.
[11] S. T. Picraux, Nucl. Inst. Meth. **182-183** (1981) 413; H. D.Carstanjen. Phys. Status. Solidi **A59** (1980) 11
[12] M. E. Stoll & T. J. Majors, Phys. Rev. **B24** (1981) 2859 The susceptibility of NbH$_x$ has been measured by S. Aronson et al, J. Less Common Met. **21** (1970) 439
[13] W. Potzel et al, Hyp. Int. **72** (1992) 191
[14] D. H. W. Carstens & P. D. Encinias, Los Alamos Natl. Lab. Report LA-UR-90-1762 (1990).
[15] G.C. Abell & D.F. Cowgill, Phys. Rev. **B44** (1991)4178
[16] G. C. Abell & S. Attalla, Phys. Rev. Lett. **59** (1987) 995
[17] LSND Collaboration, Phys. Rev.Lett. **79** (1998) 1774
[18] KamLand Collaboration, Phys. Rev. Lett. **94** (2005) 081801